%% file: main.tex
\documentclass[10pt,conference]{IEEEtran}
\usepackage{cite}
\usepackage{amsmath,amssymb,amsfonts}
\usepackage{algorithmic}
\usepackage{graphicx}
\usepackage{textcomp}
\usepackage{xcolor}
\usepackage{multirow}
\usepackage{subfigure}
\usepackage{tcolorbox}
\usepackage{enumitem}
\usepackage{array}

\newcommand{\blank}[1]{\hspace*{#1}\linebreak[0]}

\begin{document}

\title{The Impacts of Sentiments and Tones in Community-Generated Issue Discussions}

\author{\IEEEauthorblockN{Arghavan Sanei, Jinghui Cheng}
\IEEEauthorblockA{\textit{Department of Computer and Software Engineering} \\
\textit{Polytechnique Montreal, QC, Canada}\\
arghavan.sanei@polymtl.ca, jinghui.cheng@polymtl.ca}
\and
\IEEEauthorblockN{Bram Adams}
\IEEEauthorblockA{\textit{School of Computing} \\
\textit{Queen's University, ON, Canada}\\
bram.adams@queensu.ca}
}

\maketitle
\begin{abstract}
The diverse community members who contribute to the discussions on issue tracking systems of open-source software projects often exhibit complex affective states such as sentiments and tones. These affective states can significantly influence the effectiveness of the issue discussions in elaborating the initial ideas into actionable tasks that the development teams need to address. In this paper, we present an extended empirical study to investigate the impacts of sentiments and tones in community-generated issue discussions. We created and validated a large dataset of sentiments and tones in the issues posts and comments created by diverse community members in three popular open source projects. Our analysis results drew a complex picture of the relationships between, on the one hand, the sentiments and tones in the issue discussions, and on the other hand, various discussion and development-related measures such as the discussion length and the issue resolution time. We also found that when factors such as the issue poster roles and the issue types were controlled, sentiments and tones had varied associations with the measures. Insights gained from these findings can support open source community members in making and moderating effective issue discussions and guide the design of tools to better support community engagement.
\end{abstract}


\input{s_introduction}
\input{s_relatedwork}
\input{s_methods}
\input{s_results}
\input{s_discussion}
\input{s_conclusion}

\bibliographystyle{IEEEtran}
\bibliography{references}

\end{document}

%% file: s_introduction.tex
\section{Introduction}
The maintenance and evolution of many open-source software (OSS) projects rely on the issues generated by the members of the community formed around the project~\cite{Heck2013}. Modern software engineering tools such as \textit{Issue Tracking Systems} (ITSs) enabled the collection and elaboration of these issues on a large scale. Many ITSs (e.g., Jira and GitHub Issues) do not only allow community members to manage bug reports, feature requests, enhancements, questions, and other related topics about the project, but they also provide communication and community building features to facilitate the discussion about these issues~\cite{Arya2019}. Thus, an effective discussion plays a crucial role in the development of the initial ideas posted in the issues into elaborated and actionable tasks for the development team to address.

The community members who contribute to the creation and discussion of these issues usually have a diverse background, serving different roles in the project, and thus having varied values and perspectives~\cite{Cheng2019ChaseOSS,Trinkenreich2020HiddenFigures}. On the one hand, OSS communities' heterogeneous nature provides diverse perspectives that help a healthy advancement of the project. On the other hand, consolidating the varied perspectives in a large amount of community-generated requests and discussions is a challenging issue~\cite{Wang2020}. This challenge is particularly exacerbated by the complex emotions and affective states from the community participants manifested in the issues~\cite{Ortu2016}.

The importance of such affective states embedded in issue posts and discussions is widely recognized to affect the productivity of the engineers and the health of the project team~\cite{Mantyla2016Burnout}. These affective aspects also contain useful information that can be exploited by the software development teams (e.g., for predicting the time required to complete a code review~\cite{Ikram2019}). However, a lot of recent research is only focused on advancing automatic emotion detection techniques (e.g., ~\cite{Novielli2020MSR,Zhang2020ICSME,Biswas2020ICSME}), rather than examining the potential use cases of these techniques. The existing studies on the impacts of emotions and affective states in the issue discussions are also more concentrated on the development aspects (e.g., bug resolution time and commit size), but not on social and community aspects (e.g., characteristics in the follow-up discussion and responses). Moreover, the emotions and affective states expressed in the issue discussions are often intertwined with other factors such as the nature and complexity of the issue to impact both social and developmental aspects of OSS projects. Thus, an extensive image is still to be painted for assessing the importance of these factors on the manner in which the OSS community members react to and address the issues.

In this paper, we aim to fill this gap through an extended empirical study to investigate the correlational impacts of sentiments and tones in the community-generated issue discussions. The general affective computing research has evolved around two major perspectives about human emotions~\cite{picard2000, scherer2000}: a dimensional perspective, which defines these factors according to one or a few spectrums or dimensions; and a categorical perspective, which considered affective states as discrete and distinguished categories. Based on these two major perspectives, we focused on two concrete concepts as frameworks to capture the affective characteristics in issue discussions: \textit{sentiments} that capture the dimensional attitude (i.e., good or bad, like or dislike) of an author expressed in a natural language statement~\cite{Pang2008,Cambria2017}; and \textit{tones} that represent categorical language and speech features that indicate the mental or emotional states of the author~\cite{Yin2017}. Leveraging these two constructs, we pose the following research questions:

\begin{enumerate}[noitemsep, leftmargin=*, itemindent=18pt, label={\textbf{RQ{{\arabic*}}}:}]
    \item What are the descriptive characteristics of sentiments and tones in OSS issue posts and their comments?
    \item What are the relationships of sentiments and tones in the issue posts and sentiments and tones in their comments?
    \item How are the sentiments and tones of the issue posts and their comments associated with issue discussion features? We particularly ask:
        \begin{enumerate}[noitemsep, leftmargin=*, wide=0pt, label={\textit{RQ3.{{\arabic*}}}:}]
            \item How are the sentiments and tones of the issue posts associated with the time until the first response to the issue?
            \item How are the sentiments and tones of the issue posts associated with the length of the issue discussion?
            \item How are the sentiments and tones of the issue comments associated with the length of the issue discussion?
        \end{enumerate}
    \item How are the sentiments and tones of the issue posts and their comments associated with the issue resolution time?
\end{enumerate}

In order to answer these research questions, we created, validated, and analyzed a large dataset comprising the sentiments and tones of 32,764 issues and 192,956 comments made by diverse community members in three OSS projects. Moreover, to disentangle the impacts of the sentiments and tones in issues with different characteristics, we controlled four factors in our analysis when answering each research question: (1) the length of the issue discussion, (2) the size of the commits associated with the issue, (3) the role of issue reporter, and (4) the type of the issue. Overall, our study provides extended insights on the relationships between the sentiments and tones of issue discussions and the various factors that they might have influenced. This knowledge can have important practical implications for contributors to OSS projects and designers of tools for OSS development.

%% file: s_relatedwork.tex
\section{Related Work}
%This work is closely related to previous studies that focus on (1) analysis of OSS issue discussions, (2) classification of emotions based on psychological theories, (3) automated detection of emotions in textual data, and (4) software engineering studies about impacts of emotions. 
%We briefly review the previous studies related to our work in the following categories.

\subsection{OSS issue discussions}
Issue tracking systems (ITSs) serve as an important tool for OSS communities to complete various software engineering activities. Researchers have investigated methods and techniques that leverage the information presented in the ITSs to resolve a wide range of software engineering challenges, including requirements analysis~\cite{Heck2017,Morales-Ramirez2018}, bug triaging~\cite{Xia2017}, design rationale retrieval~\cite{Viviani2018}, defect type prediction~\cite{Patil2020}, and impact analysis~\cite{Huang2017}, to name a few. %For examle, Xia et al,~\cite{Xia2017} proposed a topic modeling algorithm that transforms the issue posts into a set of topics (i.e. clusters of terms), which are then combined with other issue features (e.g. component or product indicated in the issue report) to predict the most appropriate fixer of the issue.

In addition to leveraging the information included in the issue post, researchers have also dig into the discussion threads of the issues. These studies are based on a common understanding that the comments to the issues, made by various software stakeholders to respond to the original issue post, include rich information useful for the software development team. Bertram et al. \cite{Bertram2010} investigated the use of ITSs in collocated and closed software development teams and identified the role of the issue discussion threads as conversations that serve as ``a focal point for communication and coordination.'' More recently, Arya et al.~\cite{Arya2019} have identified 16 types of information that are commonly embedded in issue discussion threads. %, such as discussions about the solutions of the issues, inquiries or reports about the task progress, as well as suggestions of temporary workarounds to overcome the issue. 
Rath and M\:ader ~\cite{Rath2020} also identified three patterns of conversation (i.e., monolog, feedback, and collaboration) when stakeholders discuss the various topics in ITSs. These studies further demonstrated the complexity of issue discussions in the OSS context. Recognizing this complexity, Wang et al.~\cite{Wang2020} proposed the use of an argumentation model, as well as the corresponding automated technique, to help extract and consolidate the diverse topics and opinions that the OSS community members post in ITSs. Our work is built upon these previous studies to further investigate the roles of emotions in the already complex OSS issue discussions.

\subsection{Classification of human emotions}
Psychologists and affective computing researchers have investigated human emotions for decades, from which two major perspectives have emerged~\cite{picard2000, scherer2000}. The first perspective considers the emotions as a phenomenon that manifests in a few spectrums or \textit{\textbf{dimensions}}~\cite{schlosberg1954}. A widely accepted dimensional model about emotions contains two major spectrums: (1) valence (often also called sentiment), which indicates the degree of positive or negative affectivity; and (2) arousal, which denotes the intensity or strength of the emotion~\cite{picard2000}. Psychological studies have demonstrated the impact of these emotional dimensions in various cognitive functions. For example, in a recent study, Megalakaki et al. have found that texts with positive sentiment have led to better comprehension and memorization of text than those with negative valence~\cite{Megalakaki2019}. These studies motivated us to investigate the impacts of the valence, or sentiment polarities, in the OSS context.

The second major perspective for classifying emotions projects this construct onto several discrete \textit{\textbf{categories}}. The seminal work of Ekman et al., through studies about human facial expressions, proposed that emotions are comprised of six basic elements: anger, disgust, fear, happiness, sadness, and surprise~\cite{EKMAN1972}. These basic emotions serve as building blocks that can be combined to form more complex emotions, such as guilt, frustration, and love. The principle that emotions are comprised of basic elements is widely accepted in later studies. However, there is no consensus as to the number or the categories of the elementary emotional components~\cite{Jack2014, plutchik1991}. Researchers tend to agree, instead, that the complex or compound emotions reflect higher cognitive needs of humans in situations that can have more practical implications~\cite{Keith1987}. In this study, we leverage an emotion categorization framework proposed by Yin et al.~\cite{Yin2017}, which is focused on customer service-related texts. Based on a factor analysis of customer service conversations on Twitter, annotated for 53 emotional categories through a crowd-sourcing approach, they identified eight emotional factors: \textit{anxious}, \textit{excited}, \textit{frustrated}, \textit{impolite}, \textit{polite}, \textit{sad}, \textit{satisfied}, and \textit{sympathetic}. In the context our study, these emotional factors can well represent the nature of conversational problem-solving in issue discussions and match our interests in the analysis.

\subsection{Automated detection of sentiments and tones in text}
Researchers have long investigated automated techniques to detect the emotions aligned with the above two perspectives in textual data within and beyond the software engineering domain~\cite{Sailunaz2018, Novielli2019}. Various methods have been developed to identified sentiment polarities~\cite{Pang2008}. For example, SentiStrength, proposed by Thelwall et al.~\cite{Thelwall2012, Thelwall2017}, is widely used for short social media texts; it leverages a lexical approach that calculates the sentiment of the text using a predetermined dictionary of sentiment-related terms that reflect the characteristics of short web texts. Leveraging SentiStrength, Islam and Zibran~\cite{Islam2018} developed SentiStrength-SE, a tool that incorporates a dictionary and heuristics created specific for the software engineering domain. Calefato et al.~\cite{Calefato2018} have also developed a technique for sentiment analysis in the software engineering domain, called Senti4SD. This technique leverages a rich feature set that included sentiment lexicon-based, keyword-based, and semantic features. Using Support Vector Machine (SVM), Senti4SD is trained and validated using 4,423 Stack Overflow posts, manually annotated for sentiment polarity. Inspired by the recent advances in deep learning, Biswas et al.~\cite{Biswas2020ICSME} and Zhang et al.~\cite{Zhang2020ICSME} also investigated the application of pre-trained language models, BERT in particular, in sentiment classification; they found that the BERT sentiment classifier out-performed several other state-of-the-art sentiment analysis techniques on a Stack Overflow dataset. A recent replication study found that Senti4SD has achieved similar performance than BERT-based techniques; the former also out-performed the latter on a GitHub discussion dataset~\cite{novielli2020assessment}.

Because of Senti4SD's superior performance in classifying software-specific natural language data, particularly on issue discussions, we chose this technique to identify the sentiment polarity of the OSS issue posts and comments. The pre-trained Senti4SD model classifies an input natural language text into three emotion polarity categories: (1) \textbf{positive} sentiment that indicates emotions related to joy or love, (2) \textbf{negative} sentiment that connotes anger, sadness, or fear, and (3) \textbf{neutral} indicating neither positive nor negative.

Automated detection of \textit{categories} of emotions from textual data is also an active research direction. For example, Abdul-Mageed and Ungar~\cite{AbdulMageed2017} have developed a large-scale dataset and a deep learning algorithm to identify the 24 fine-grained categories of emotions in Robert Plutchik’s framework~\cite{plutchik1991} from twitter texts. Yin et al.~\cite{Yin2017} also proposed an unsupervised method that leverages the Latent Dirichlet Allocation (LDA) model to detecting \textit{tones} in customer service conversations on Twitter. Yin et al.'s work is implemented in IBM Watson Tone Analyzer. This service provides a REST API using a pre-trained machine learning model to classify natural language text inputs with the seven affective factors (i.e., tones): \textbf{excited}, \textbf{frustrated}, \textbf{impolite}, \textbf{polite}, \textbf{sad}, \textbf{satisfied}, and \textbf{sympathetic}. The service can return multiple tones with a given input and provides a confidence score (range from 0.51 to 1) for each tone it classifies. To our best knowledge, the detection of tones have not been systematically examined in the software engineering context. In a preliminary analysis, we found that the affective categories expressed in issue discussions are similar to the tones categorized by Yin et al. In our study, we directly used the IBM Watson Tone Analyzer for categorizing the tones of issue discussions.

\subsection{Impacts of emotions and affective states in software development}
%Recent studies indicate that there is increasing interest in identifying affective computing in different types of texts such as social media ~\cite{Hu2018, Thelwall2012}, reviews ~\cite{alshayban2020, Ikram2019}, issue discussion ~\cite{cheruvelil2019}, developers' emotion ~\cite{girardi2020},  software engineering ~\cite{Novielli2019, Valdez2020}, etc.. However, our concentration in this paper is on the influences of affective computing in software engineering. 
Much of the previous work investigating the impacts of affective states in the context of software development focused on sentiment analysis. These studies are often based on the assumption that understanding the sentiments expressed in software artifacts can lead to improvements in software engineering tools and communications among the developers~\cite{Novielli2019}. The role of sentiment is investigated in commit messages, code reviews, exchanges in Q\&A websites, as well as issue discussions. For example, Sinha et al.~\cite{Sinha2016} found that while most commit messages had a neutral sentiment, negative sentiments appeared more frequently than positive ones. They further identified a strong correlation between these sentiments and the size of the commits, with positive sentiments associating with smaller commits. Besides, Huq et al.~\cite{Huq2020} found that the sentiments expressed in messages on commits that introduced or fixed bugs were more negative than those on other commits. Focusing on the role of sentiments in code review activities, Asri et al.~\cite{Ikram2019} found that reviews embodied with negative sentiments took more time to be completed than reviews with positive or neutral comments. Calefato et al.~\cite{Calefato2018Howto} also demonstrated that on Stack Overflow, the questioners embedding neutral sentiments were more likely to have a satisfiable answer.

Our work is most closely related to previous studies investigating the impacts of emotions and affective states expressed in software issue discussions. Focusing on both dimensional and categorical emotions, Ortu et al.~\cite{Ortu2015} found that positive emotions were correlated with shorter issue fixing time, based on an analysis of a Jira dataset. Focusing on security-related discussions on GitHub, Pletea et al.~\cite{Pletea2014} also found that sentiments in the discussion of this type of issues are more negative than other types. More recently, Valdez et al.~\cite{Valdez2020} found that comments in unresolved Jira issues tend to express more negative sentiments than in resolved issues. Our work builds on these previous studies and aims for an extended analysis of the impacts of sentiments and tones in OSS issue discussions.

%% file: s_methods.tex
\section{Methods}
\subsection{Data Selection}
To answer our research questions, we focused on collecting and analyzing data from three popular open-source Machine Learning (ML) libraries: (1)  TensorFlow\footnote{https://github.com/tensorflow/tensorflow}, developed and maintained by the Google Brain team, (2) Scikit Learn\footnote{https://github.com/scikit-learn/scikit-learn}, a community-driven project that is maintained by a team of volunteers, and (3) PyTorch\footnote{https://github.com/pytorch/pytorch}, developed by Facebook's AI Research lab (FAIR). We focus on these projects because of the following reasons. First, these well-known open-source ML open source projects have attracted a diverse and active community. Because they successfully focus on the ease-of-use in building ML models, algorithms, and applications, the users of these libraries do not only include computer scientists and ML experts, but they also extend to domain experts, developers, students, and amateurs who use or are interested in various ML techniques. This heterogeneous user base can help provide rich and diverse information about the impacts of sentiments and tones in the issue discussion systems. Second, these ML libraries are consist of a rapidly increasing number of issues, comments, and commits since users with diverse backgrounds and perspectives have widely used and contributed to them. These artifacts are also well managed. For example, among the more than 34,500 closed issues in the three projects, around 56\% were labeled and 95\% included at least one comment. Thus these projects provide us a large, heterogeneous, and high-quality dataset. Finally, all authors of this paper are familiar with these ML libraries, making it possible to perform manual analysis and sanity checks of the automated techniques and appropriately interpret the analysis results.

\subsection{Data collection}
We focused on collecting closed issues that had at least one comment, as well as the associated comments, commits, and meta-data on the three GitHub repositories. We chose to target the analysis on the closed issues because they represent the entire life-cycle of issue discussions with embedded sentiments and tones; we only included issues that have at least one comment because we aim to understand the impacts of sentiments and tones on the issue posts to those in its comments. While issues can still receive comments even after they are closed, we did not differentiate them from the other comments. Collection of data from the three GitHub repositories was performed in March 2020, using the GitHub REST API\footnote{https://developer.github.com/v3/}. Table~\ref{tab:data_total} summarizes the numbers of issues, comments, and commits in our dataset.

\begin{table}[ht]
\vspace{-6pt}
\centering
\caption{Summary of our dataset}
\begin{tabular}{lccc}
\hline
\multicolumn{1}{c}{Project} &
\multicolumn{1}{c}{Issues} &
\multicolumn{1}{c}{Comments} &
\multicolumn{1}{c}{Issues linked to a commit} \\ \hline
Scikit Learn & 5,456 %5,701%
& 35,402 & 539 \\
 PyTorch  &  7,934 %8,803%
 & 36,048 & 1,955  \\
Tensorflow   &  19,374 %20,068%
& 121,506 & 1,144 \\
\hline
Total & 32,764 & 192,956 & 3,638\\
\hline
\end{tabular}
% \vspace{-6pt}
\label{tab:data_total}
\end{table}

\subsection{Initial data pre-processing and analysis}
For all natural language data, including the issue posts and comments, we removed and replaced the textual contents based on the criteria provided in Table~\ref{tab:preprocess} in order to remove noises in the text to ensure correct sentiment and tone detection. Stop words were not removed in consistency with previous research in sentiment analysis~\cite{Calefato2018}. We also did not do any tokenization, lemmatization, or stemming in data pre-processing since the analysis tools we use (Senti4SD and IBM Tone Analyser) perform these tasks themselves. Once these preliminary data pre-processing steps were done, we fed the data into the two sentiment and tone analysis tools to obtain the initial identification results. 

\begin{table}[ht]
\centering
\caption{Textual contents removed during pre-processing.}
\begin{tabular}{ll}
\hline
\multicolumn{1}{c}{\textbf{Content Removed}} & \multicolumn{1}{c}{\textbf{RegEx used to detect the content}} \\ \hline
Code snippets & \verb|```|, \verb|```python| \\
HTML tags & \verb|<!--.*?-->|, \verb|<[^>]*>| %<[^>]*>%
\\
URLs & \verb|https/S+|, \verb|http/S+| \\
Reference quotation & \verb|>/S+|, \verb|<.*?>|\\
System path & \verb|c:/S+|, \verb|e:/S+|, etc. \\
Mentioning someone & \verb|(@[A-Za-z0-9]+)| \\
Issue templates & \verb|### next steps|, \verb|** logs|, etc. \\
% Description comment\\ in issue \jc{Reference quotation or HTML comment?} & '\textless!--.*?--\textgreater' \\
%removing numbers & '[0-9]+' \\
\hline
\end{tabular}
\vspace{-12pt}
\label{tab:preprocess}
\end{table}

\subsection{Investigating the performance of the analysis tools}
If not used properly, sentiment and tone analysis tools are known to be unreliable~\cite{Imtiaz2018,Novielli2020MSR}. In order to achieve a satisfactory level of reliability and ensure the performance of the analysis tools we use, we performed the following investigation steps.

\subsubsection{Gold standard}
We performed manual annotation of a sample of the dataset to obtain a gold standard of sentiment and tone identification. To ensure diversity and to capture all sentiment and tone categories in the gold standard, we conducted a Multivariate Stratified Sampling~\cite{Barcaroli2014} based on the initial identification of the tools. Particularly, we sampled 200 issues and 400 comments across the three repositories from our dataset for manual analysis. Two researchers then independently coded the sample on dimensional sentiment (i.e., negative, neutral, or positive) and categorical tones (i.e., excited, frustrated, impolite, polite, sad, satisfied, and sympathetic); initial agreement based on Cohen's Kappa was considered `substantial' ($\kappa=0.66$ and $0.62$ for sentiments and tones, respectively)~\cite{Viera2005Kappa}. The two coders then conducted meetings of more than six hours in total to discuss their coding and reached a full agreement, which is then considered as the gold standard of analysis. We used Cohen's Kappa as a measure to evaluate the agreement between the automated identification results and the gold standard~\cite{Viera2005Kappa}.

\subsubsection{Investigating additional data pre-processing steps}
With the support of the gold standard, we investigated the effects of the following data pre-processing steps on the performance of the sentiment and tone identification tools on the sample: (1) removing emojis and emoticons, (2) removing punctuation, (3) removing numbers, (4) removing the name of Python libraries, classes and methods, and (5) changing all letters to lowercase. All five additional pre-processing steps decreased the $\kappa$ agreement values between the tool results and the gold standard. Consequently, we decided to omit these steps.% and never considered them in our pre-processing data. The considered removed contents are mentioned in Table ~\ref{tab:preprocess}.Besides, one of the goal in selecting these 5 experiments is checking how these elements impact on the tools' classification and how much they increase or decrease the agreement between gold standard and tools. %The overall analysis results had an average Kappa of 0.56 for sentiment and 0.39 for tones.

\subsubsection{Investigating the granularity of analysis}
In order to achieve more accurate identification of the overall sentiments and tones in issue posts and comments, we investigated the tools' performance when the identification is conducted on two different levels of granularity: (1) the post level, in which the entire post of issue discussion is fed into the tools for sentiment and tone identification; and (2) the sentence level, in which the issue discussion posts were separated into sentences, the sentiments and tones of each sentence are identified with the tools, and then the results aggregated to the post level to represent the overall sentiments and tones. To aggregate the sentence-level sentiments into the post level, we assigned a numerical value to each sentiment category (i.e., positive as 1, neutral as 0, and negative as -1) and summed up the sentiments identified in each sentence of a post; then, based on the sign of this summed number, we considered the entire post as being positive (sum~\textgreater~0), negative (sum~\textless~0) or neutral (sum~=~0). For aggregating the tones, we first summed up the confidence level provided by IBM Tone Analyzer for each type of tone in each sentence of a post; then, we selected up to three tones that had the highest overall confidence level to represent the overall tones of the post. These aggregation methods were informed by the insights we obtained through the manual annotation process about how humans identify sentiments and tones in a multi-sentence text.

Results on the sampled 600 posts of issue discussions (including 200 issue posts and 400 comments across the three repositories) show that the sentence-level sentiment and tone analysis resulted in higher accuracy (in average, $\kappa=0.77$ for sentiment and $0.63$ for tone) than the post-level analysis(in average, $\kappa=0.66$ for sentiment and $0.62$ for tone). As a result, we used the sentence-level analysis with the aforementioned aggregation methods to identify the sentiments and tones of each issue description and comment in our dataset. Overall, the above investigation helps us determine an optimal pre-processing and configuration strategy for the automated tools.

\subsection{Sentiment and tone identification}
We applied the aforementioned data pre-processing steps and the sentence-level analysis on the remaining issue posts collected from the three repositories, then used Senti4SD and IBM Watson Tone Analyzer to identify the sentiments and tones in the entire dataset. To further assess the accuracy of the automatically identified sentiments and tones, we again sampled another test set of 100 issue posts and 200 issue comments and performed the same manual analysis as we did when creating the gold standard. The agreements between the automated methods and our manual annotation results on this test set were ``substantial'' or ``almost perfect'' (in average, $\kappa=0.81$ for sentiments and $0.69$ for tones~\cite{Viera2005Kappa}), indicating that we could be confident about the automatically identified sentiments and tones for further analysis.

In the data analysis, we directly used the automated detection results for the sentiments and tones of the issue posts. To identify the overall sentiments and tones presented in all the comments to an issue, we further aggregated the automatic detection results in each comment on an issue. We chose the most frequently appeared sentiment among all comments of an issue to represent the overall sentiment. For the dominant tone, we calculated the sum of each tone's confidence scores in all the comments on an issue; we then chose the tone with the highest accumulated confidence score as the overall tone. We then answered the research questions based on these data.

\subsection{Data analysis}

Our dataset, data analysis code, and the full analysis results can be found at https://doi.org/10.5281/zenodo.4486943. To answer RQ1, we calculated descriptive statistics of the identified sentiments and tones in the issue discussions of the three projects. We then analyzed the impacts of the sentiments and tones embedded in the issue discussions through the following approach. To answer RQ2, we considered the sentiments and the tones of the issue posts as two independent variables and the aggregated sentiments and the tones of the issue comments as dependent variables. Then we used the chi-squared tests to examine the correlations among the independent and dependent variables. When the test indicates a significant correlation, we used Cramer's V to evaluate the strength of the correlation. Cramer's V ranges between 0 to 1; a value of greater than 0.15 can be considered to indicate a \textit{strong} correlation, while a value of greater than 0.25 indicates a \textit{very strong} correlation~\cite{Akoglu2018Cor}.

To answer RQ3 and RQ4, we considered four independent variables: (1) issue post sentiment, (2) issue post tone, (3) overall sentiment of issue comments, and (4) overall tone of issue comments. The dependent variables in RQ3 are (1) the time between issue post and the first comment (RQ3.1) and (2) the number of comments included in the issue (RQ3.2); the dependent variables in RQ4 are (1) the times between the issue post and the first commit and (2) the time between issue post and the last commit. Because we cannot assume a normal distribution of the data, we used the non-parametric Kruskal-Wallis tests to evaluate the association between the independent and the dependent variables. Due to the large sample size in our dataset (thus a strong statistical power), we considered a low alpha level of .001 to avoid making Type I errors. When a statistical significant association is found between an independent and a dependent variable, we then conducted post-hoc analyses on each value pair of the independent variable using the Mann–Whitney U test with Bonferroni correction to understand the groups that contributed to the association. Bonferroni correction was used to mitigate the issue of multiple comparisons. The alpha levels were set at .01 in these tests.

Sentiments and tones in issues with different characteristics may have different impacts. As a result, we also aimed to answer RQ2, RQ3, and RQ4 on the subset of the issues that had varied (1) discussion lengths, (2) sizes of commit, (3) poster roles, and (4) types. For the issue \textit{\textbf{discussions length}}, we defined three categories depending on the number of comments of the issues. The thresholds were set to be the 50th and the 99th percentile of the distribution of the number of comments. Thus, we defined \textit{short} issues to have [1,~4) comments, \textit{medium} issues to have [4,~34) comments, and \textit{long} issues to have 34 or more comments. For the \textit{\textbf{size of commit}} made to resolve the issue, we again defined three categories separated by the 25th and 50th percentile of the distribution of the total lines of change in the commits associated with the issues. We defined issues with \textit{small} commits to have a total of [1,~8) lines, issues with \textit{medium} commits to have a total of [8,~117) lines, and issues with \textit{large} commits to have 117 or more lines. These thresholds were set based on our analysis of the data distribution and were inspired by~\cite{Sinha2016}. In our preliminary analysis, the thresholds in individual projects are very similar. As a result, the categories and the analysis were conducted on the combined data of the three repositories. For the \textit{\textbf{roles of the issue poster}}, we directly used the roles identified in the GitHub repository to include \textit{contributor} (i.e. individuals who have contributed to the codebase) and \textit{non-contributor} (i.e. individuals who have never committed to the codebase). For the \textbf{\textit{types of the issues}}, we aimed to separate \textit{bug reports} and \textit{feature requests} based on the labels used in the issue tracking system on GitHub. All three projects have extensively labeled the issues. We used the labels such as `type: bug', `topic: crash', and `regression' as indicators of bug reports; labels such as `feature', `proposal', `enhancement' were used to identify feature reports. The full list of labels we used can be found in our data analysis code.

%% file: s_results.tex
\section{Results}
\subsection{Descriptive results}
\begin{figure} [t]
\centering     %%% not \center
\includegraphics[width=\linewidth]{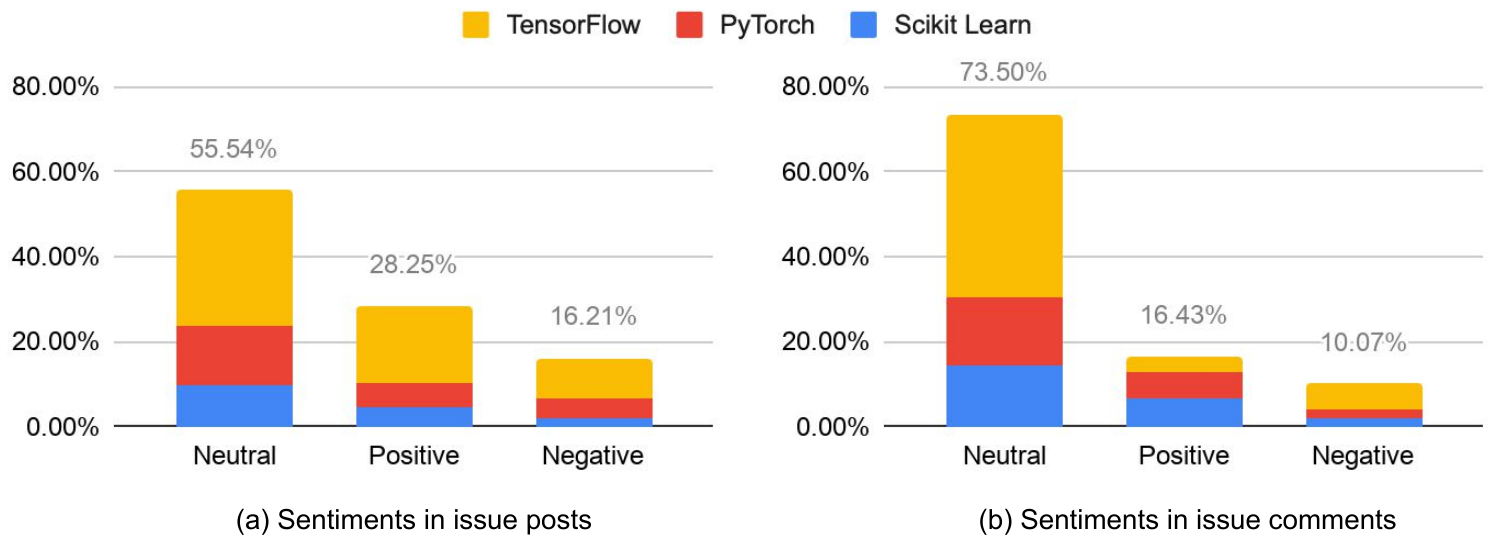}
\vspace{-20pt}
\caption{Frequency of sentiments in issue posts and comments}
% \vspace{-6pt}
\label{fig:sentiment_frequency}
\end{figure}

\begin{figure} [t]
\centering     %%% not \center
\includegraphics[width=\linewidth]{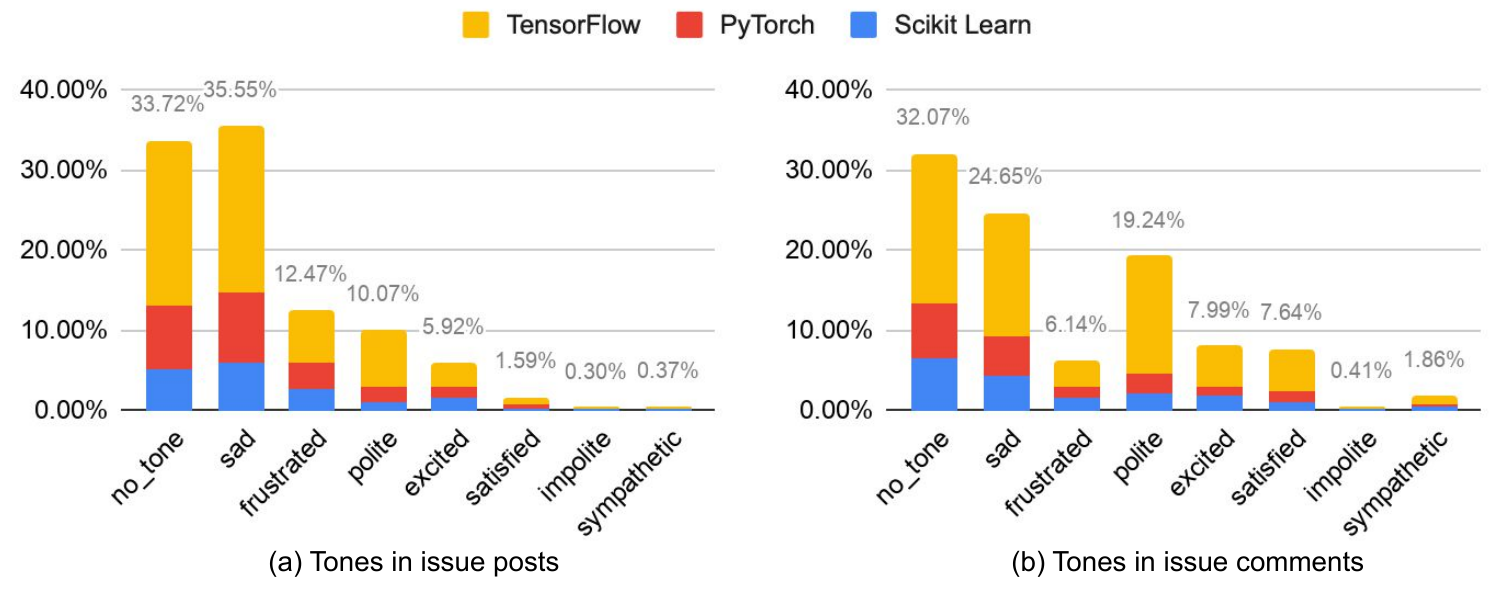}
\vspace{-20pt}
\caption{Frequency of tones in issue posts and comments}
% \vspace{-12pt}
\label{fig:tone_frequency}
\end{figure}

In our dataset, the most prominent sentiment among the issue posts and comments was \textit{neutral}, followed by \textit{positive}. Other than \textit{no tone}, the most frequent tone in issue posts was \textit{sad}, followed by \textit{frustrated} and \textit{polite}. In comments, the most frequent tone was also \textit{sad}, followed by \textit{polite} and \textit{excited}. Because we only had very few data points on \textit{impolite} and \textit{sympathetic} tones, we excluded these two tones for further analyses. Figures~\ref{fig:sentiment_frequency} and \ref{fig:tone_frequency} summarize the frequency of the identified sentiments and tones.

\vspace{4pt}
\begin{tcolorbox}[colframe=black,colback=gray!10,boxrule=0.5pt,arc=.3em,boxsep=-1mm]
\textbf{Insight 1}: Other than the \textit{neutral} sentiment and \textit{no tone}, the \textit{positive} sentiment and the \textit{sad} tone most frequently appeared in our dataset of issue discussions.
\end{tcolorbox}

\begin{table}[t]
\caption{Correlation between sentiments and tones on the same text}
\begin{tabular}{llll}
\hline
Project & Relationship & Cramer's V & Strength~\cite{Akoglu2018Cor} \\\hline
\multirow{2}{*}{PyTorch} & Issue: sentiment-tone & 0.334** & Very strong \\
 & Comment: sentiment-tone & 0.239** & Strong \\
\hline
\multirow{2}{*}{TensorFlow} & Issue: sentiment-tone & 0.385** & Very strong \\
 & Comment: sentiment-tone & 0.303** & Very strong \\
\hline
\multirow{2}{*}{Scikit learn} & Issue: sentiment-tone & 0.347** & Very strong \\
 & Comment: sentiment-tone & 0.193** & Strong \\
\hline
\multirow{2}{*}{All} & Issue: sentiment-tone & 0.362** & Very strong \\
 & Comment: sentiment-tone & 0.286** & Very strong\\
\hline
\end{tabular}\vspace{2pt}\\
\raggedleft** indicates $p<0.001$ with chi-squared test
% \vspace{-6pt}
\label{tab:sent-tone-overall}
\end{table}

\begin{table}[t]
\caption{Correlation between sentiments and tones in the issue posts and those in their comments of the TensorFlow project}
\begin{tabular}{ll|cc}
\hline
 &  & \multicolumn{2}{c}{\textbf{Issue post}} \\
 &  & Sentiment & Tone \\\hline
\multirow{2}{*}{\textbf{Comments}} & Sentiment & Cramer's V=0.054** & Cramer's V=0.059** \\
 & Tone & Cramer's V=0.055** & Cramer's V=0.035**\\\hline
\end{tabular}\vspace{2pt}\\
\raggedleft** indicates $p<0.001$ with chi-squared test
% \vspace{-12pt}
\label{tab:issue-comment-tensorflow}
\end{table}

\subsection{Correlations among sentiments and tones}
Using chi-squared tests of independence, we have found significant associations among the sentiments and tones of the same text (either issue posts or comments) in all three projects with strong or very strong relationships indicated by Cramer's~V~\cite{Akoglu2018Cor} (see Table~\ref{tab:sent-tone-overall}). When issues are separated based on the discussion length, commit size, poster role, or type of issue, the correlations in all categories remain significant with strong coefficients. These results further indicated the reliability of the tools we used in identifying sentiments and tones of issue posts and comments.

To answer RQ2, we analyzed the relationships of sentiments and tones in the issue posts and those in the corresponding comments. When all issues are combined, in PyTorch and Scikit-learn, we did not find any significant association. In TensorFlow, although all associations were significant at the alpha level .001, the coefficients were all weak (see Table~\ref{tab:issue-comment-tensorflow}).

When issues are separated based on the four factors, the relationships of sentiments and tones between the issue posts and the corresponding comments in all separated categories remain mostly insignificant, and if significant, have a weak coefficient. The only exceptions were found when the issues led to large commits, in which the sentiments and tones of the issue posts were significantly correlated with the tones of the issue comments ($p<0.001$) with moderate or strong coefficients (Cramer's $V=0.151$ and $0.145$, respectively).

\vspace{4pt}
\begin{tcolorbox}[colframe=black,colback=gray!10,boxrule=0.5pt,arc=.3em,boxsep=-1mm]
\textbf{Insight 2}: The relationships between the sentiments and tones in issue posts and those in the corresponding comments were only found in issues that have large codebase impacts.
\end{tcolorbox}

\subsection{Relationships with discussion measures}
The following sections answer RQ3.1, RQ3.2, RQ3.3, respectively.

\subsubsection{Effects of sentiments and tones in issue posts on the time to the first comment}

\begin{table}[t]
\caption{Summary of the relationships of sentiments and tones in issue posts with the time to first comment}
\begin{tabular}{l|ll}
\noalign{\hrule height 1pt}
 & \multicolumn{1}{c}{Issue post sentiment} & \multicolumn{1}{c}{Issue post tone} \\
\noalign{\hrule height 1pt}
 All issues & neu\textgreater neg**, pos\textgreater neu** &
 \begin{tabular}[c]{@{}l@{}} exi\textgreater no*, fru\textgreater no**, \\ no\textgreater pol**, no\textgreater sad**\end{tabular}
  \\
  \hline
% \begin{tabular} \end{tabular}  
Short issues &	neu\textgreater neg**, pos\textgreater neg** &
\begin{tabular}[c]{@{}l@{}} no\textgreater pol**, no\textgreater sad** \end{tabular}  
\\
  Medium issues &	-- &
  \begin{tabular}[c]{@{}l@{}} exi\textgreater no**, no\textgreater fru**, \\ no\textgreater sad**, sad\textgreater pol*\end{tabular}
%  Long issues	&--&	-- \\
  \\
  \hline
 % small commit	&--	&-- \\
 % Medium commit &--&	-- \\
 Large commit	&--	&   \begin{tabular}[c]{@{}l@{}} no\textgreater exi**, exi\textgreater fru**,\\ pol\textgreater exi**, sad\textgreater exi** \end{tabular} \\
 
 \hline
% Contributor poster	&--	&-- \\
 Non-contributor &	neu\textgreater neg**, pos\textgreater neu** &
 \begin{tabular}[c]{@{}l@{}}
 exi\textgreater no**, no\textgreater fru**,\\ no\textgreater sad**, no\textgreater pol**
 \end{tabular}\\
 \hline
% Feature requests	&--	&-- \\
 Bug reports	& neg\textgreater neu*, pos\textgreater neu**	& no\textgreater sad**
 \\
\noalign{\hrule height 1pt}
\end{tabular}\vspace{2pt}\\
\raggedleft{Categories with no significant results using Kruskal-Wallis test are omitted.\\
* indicates $p<0.01$ and ** indicates $p<0.001$ in post-hoc analysis.\\}
\vspace{-6pt}
\label{tab:post-firstcomment-summary}
\end{table}

Table~\ref{tab:post-firstcomment-summary} summarizes the statistical significant associations in all issues and in the various issue categories we focused on. When all issues were considered, significant differences ($p<0.001$) were found on the time for someone to make the first comment among the issue posts with different sentiments and tones. The post-hoc analysis revealed that a \textit{negative} sentiment in issue posts was significantly associated with shorter time to first comment (mean = 6.55 days) than a \textit{neutral} sentiment (mean = 8.65 days); additionally, compared with \textit{no tones} in issue posts (average time to first comment is 9.53 days), \textit{excited} and \textit{frustrated} tones significantly resulted in longer time to first comment (mean = 11.95 and 9.99 days, respectively), while \textit{polite} and \textit{sad} tones in shorting time (mean = 6.91 and 6.52 days, respectively).

When the issues were separated into categories based on \textbf{discussion length}, Kruskal-Wallis tests have identified significant differences in the response time among different post sentiments only in the \textit{short} discussion category; the impacts of tones in issue posts were also only observed in \textit{short} and \textit{medium} discussion categories. For issues with different \textbf{commit sizes}, the only significant result was found in the \textit{large} commit category between the tones of issue posts and the time to the first comment. When we consider the \textbf{roles of issue posters}, a significant difference was found only in the \textit{non-contributor} group. Further, considering issues with different \textbf{types}, Kruskal-Wallis tests have identified significant differences in \textit{bug reports} only.

\vspace{4pt}
\begin{tcolorbox}[colframe=black,colback=gray!10,boxrule=0.5pt,arc=.3em,boxsep=-1mm]
\textbf{Insight 3.1}: A \textit{negative} sentiment and a \textit{sad} or \textit{polite} tone in issue posts tend to associate with shorter response time, most apparently in short discussions, issues posted by non-contributors, and bug reports. However, in issues that require substantial code change, an \textit{excited} tone in issue posts tend to associate with shorter response time.
\end{tcolorbox}

\subsubsection{Relationships between sentiments and tones in issue posts and the discussion length}

\begin{table}[t]
\caption{Summary of the relationships of sentiments and tones in issue posts with the discussion length}
\begin{tabular}{l|ll}
\noalign{\hrule height 1pt}
 & \multicolumn{1}{c}{Issue post sentiment} & \multicolumn{1}{c}{Issue post tone} \\
\noalign{\hrule height 1pt}
 All issues &  neg\textgreater neu**, neg\textgreater pos**& \begin{tabular}[c]{@{}l@{}}	sad\textgreater no**, sad\textgreater pol**,\\ sad\textgreater sat** \end{tabular}\\
 \hline
 Medium issues	& neg\textgreater neu**, pos\textgreater neu**	& pol\textgreater no*, sad\textgreater no** \\

  \hline
  Small commit	& neg\textgreater neu*, pos\textgreater neu**&-- \\
  Medium commit &neg\textgreater neu**, pos\textgreater neu* &	sad\textgreater no** \\
 Large commit	& neg\textgreater neu**,pos\textgreater neu**	&   \begin{tabular}[c]{@{}l@{}}no\textgreater exi**, fru\textgreater exi**, \\ pol\textgreater exi**, sad\textgreater exi** \end{tabular} \\
 
 \hline
 Contributor	& neg\textgreater neu*, pos\textgreater neu** &	sad\textgreater no** \\
 Non-contributor &	
 neg\textgreater neu**, neg\textgreater pos**	&
 \begin{tabular}[c]{@{}l@{}}
 sad\textgreater no**, sad\textgreater pol**, \\ sad\textgreater sat** \end{tabular}\\
 \hline
 Feature requests	& neg\textgreater neu*, pos\textgreater neu* &
 \begin{tabular}[c]{@{}l@{}} exi\textgreater no**, pol\textgreater no**,\\ sad\textgreater no** \end{tabular} \\
 Bug reports	& pos\textgreater neu** &	--
 \\
\noalign{\hrule height 1pt}
\end{tabular}\vspace{2pt}\\
\raggedleft{Categories with no significant results using Kruskal-Wallis test are omitted.\\
* indicates $p<0.01$ and ** indicates $p<0.001$ in post-hoc analysis.\\}
\vspace{-6pt}
\label{tab:post-length-summary}
\end{table}

Table~\ref{tab:post-length-summary} summarizes the results. When all issues were included, significant differences on issue discussion length (i.e. the number of comments posted on the issue) were observed for both different sentiments and different tones in the issue posts, using Kruskal-Wallis test at alpha level .001. The post-hoc analysis revealed that a \textit{negative} sentiment in issue posts was significantly associated with longer issue discussions (mean = 6.20 comments) than \textit{neutral} issue posts (mean = 5.62 comments). Similarly, a \textit{sad} tone in issue posts were significantly associated with longer discussions (mean = 6.16 comments) than when the issue posts were dominant by \textit{satisfied} or \textit{polite} tones (mean = 4.80 comments, and 5.81 comments, respectively) or demonstrated \textit{no tone} (mean = 5.59 comments).

When considering the \textbf{issue discussion length}, the short and long groups contained too little variance on the independent variable (i.e., discussion length) to produce meaningful results. Thus, we focused the analysis on \textit{medium}-length issues (i.e., issues that have between four and 34 comments); for this group, the Kruskal-Wallis test has identified significant differences. For issues linked to commits, we also found significant differences in all three \textbf{commit size} categories. When we consider the \textbf{roles of issue posters} and different \textbf{issue types}, significant differences were also found in all categories.

\vspace{4pt}
\begin{tcolorbox}[colframe=black,colback=gray!10,boxrule=0.5pt,arc=.3em,boxsep=-1mm]
\textbf{Insight 3.2}: A \textit{negative} sentiment and a \textit{sad} tone in issue posts tend to be associated with longer discussions. In issues resulted in large code impact, an \textit{excited} tone was associated with shorter discussions.
\end{tcolorbox}

\subsubsection{Relationships between sentiments and tones in issue comments and the discussion length}
\begin{figure}[t]
\includegraphics[width=.38\linewidth]{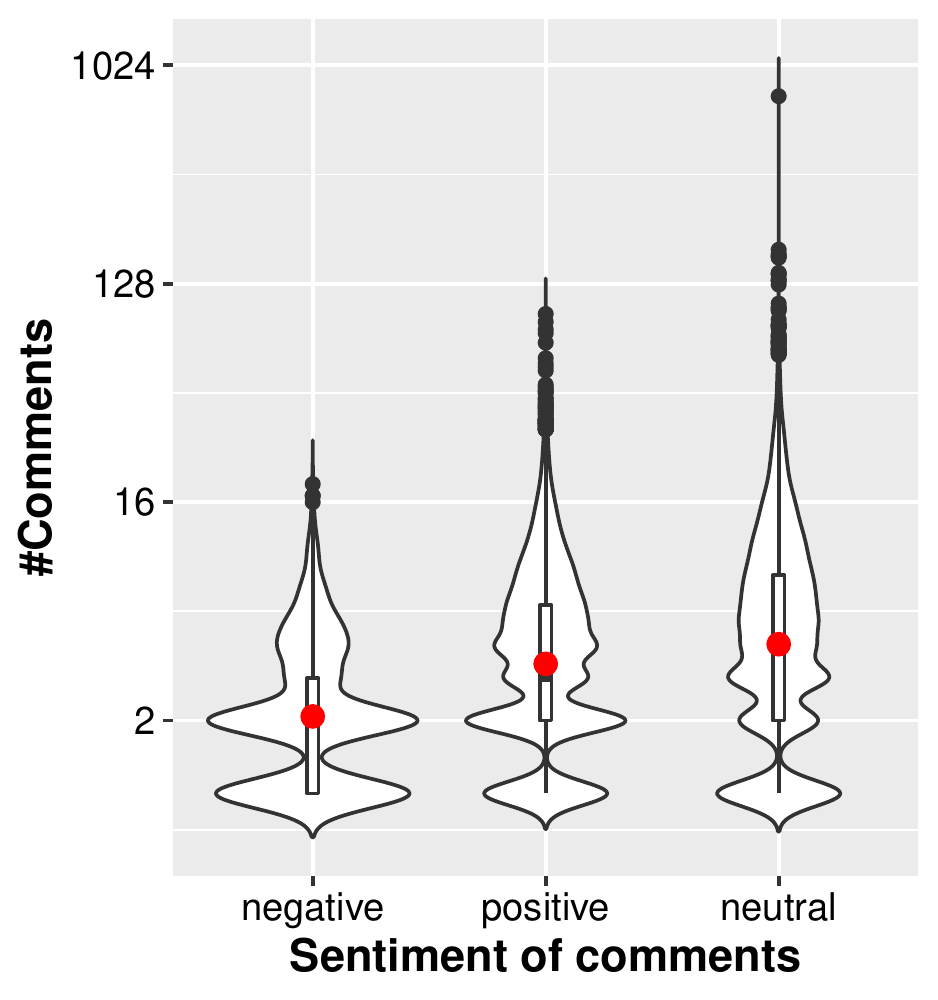}\hfill
\includegraphics[width=.6\linewidth]{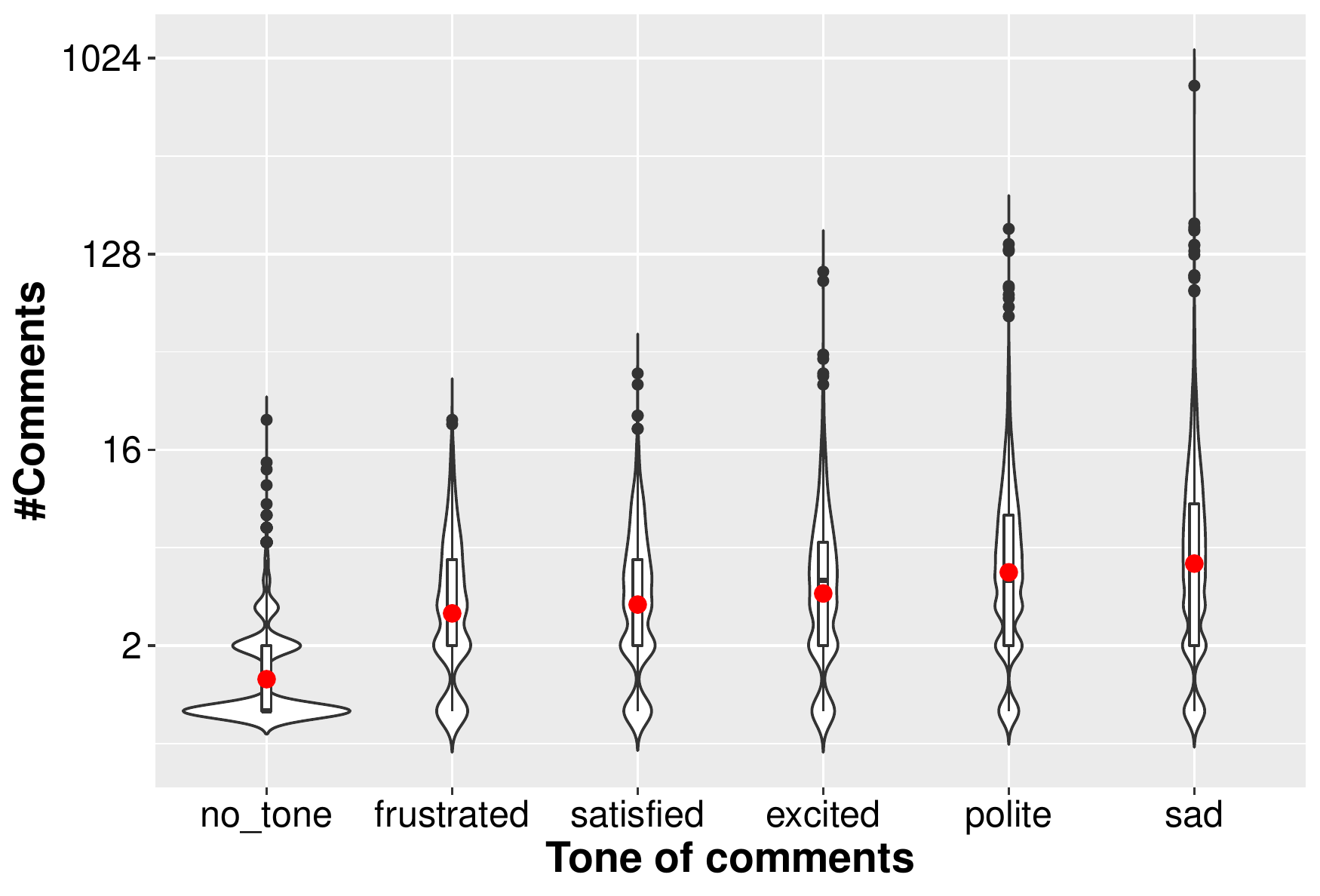}
\caption{Distributions of issue discussion lengths based on the sentiments and tones in issue comments.}
% \vspace{-12pt}
\label{fig:comment-discussion-length-with-boxplot}
\end{figure}

The Kruskal-Wallis tests have also identified significant differences on discussion length among the different overall sentiments and tones of the issue discussion comments. A post-hoc Mann-Whitney U test with Bonferroni correction has found a pair-wise difference ($p<0.001$) in which the \textit{negative} sentiment in issue discussions was associated with fewer comments (mean = 2.63 comments), followed by \textit{positive} and \textit{neutral} sentiments (mean = 4.91 comments and 6.71 comments, respectively). Similarly, the same post-hoc test also revealed that the difference on discussion length among the five analyzed tones was also pair-wise significant ($p<0.001$), in the ascending order of \textit{no tone} (mean = 1.60 comments), \textit{frustrated} (mean = 3.68 comments), \textit{satisfied} (mean = 4.04 comments), \textit{excited} (mean = 4.80 comments), \textit{polite} (mean = 6.22 comments), and \textit{sad} (mean = 7.55 comments); see Figure~\ref{fig:comment-discussion-length-with-boxplot}. When we investigate the issues with different discussion lengths, reporter roles, and types, same pair-wise relationships were found in each sub-group of issues. For issues that included commits, however, only comment tones were significantly associated with the discussion length, having the same pair-wise relationships; comment sentiments were not associated with discussion length.

\vspace{4pt}
\begin{tcolorbox}[colframe=black,colback=gray!10,boxrule=0.5pt,arc=.3em,boxsep=-1mm]
\textbf{Insight 3.3}: The overall sentiments and tones of the issue comments were directly associated with the discussion length. In the ascending order, \textit{negative}, \textit{positive}, and \textit{neutral} comment sentiments, as well as \textit{frustrated}, \textit{satisfied}, \textit{excited}, \textit{polite}, and \textit{sad} tones in issue comments tend to associate with increasingly longer discussions.
\end{tcolorbox}

\subsection{Relationships with time to fix the issue}

\begin{table*}[t]
\caption{Summary of relationships between sentiments and tones in issue discussions and time to first/last commit}
\begin{tabular}{l|ll|ll}
\noalign{\hrule height 1pt}
 & \multicolumn{1}{c}{Issue post sentiment} & \multicolumn{1}{c|}{Issue post tone} & \multicolumn{1}{c}{Issue comments sentiment} & \multicolumn{1}{c}{Issue comments tone} \\
\noalign{\hrule height 1pt}
All issues & -- & exi\textgreater{}no**, exi\textgreater{}fru*, exi\textgreater{}sad** & neu\textgreater{}pos** & sad\textgreater{}no** \\
\hline
Short issues & -- & -- & neg\textgreater{}pos*, neu\textgreater{}pos** & -- \\
Medium issues & -- & -- & neu\textgreater{}pos** & -- \\
%Long issues & -- & -- & -- & -- \\
\hline
Small commit & -- & -- & neu\textgreater{}pos** & -- \\
Medium commit & -- & -- & -- & sad\textgreater{}no**, \\
Large commit & -- & no\textgreater{}exi**, exi\textgreater{}sad**, & -- & -- \\
\hline
Contributor & -- & -- & neu\textgreater{}pos** & -- \\
Non-contributor & -- & -- & neu\textgreater{}pos** & -- \\
%\hline
%Feature requests & -- & -- & -- & -- \\
%Bug reports & -- & -- & -- & --\\
\noalign{\hrule height 1pt}
\end{tabular}\vspace{2pt}\\
\raggedleft{Times to first and last commits generated the same results. Categories with no significant results using Kruskal-Wallis test are omitted.\blank{1cm}\\
* indicates $p<0.01$ and ** indicates $p<0.001$ in post-hoc analysis.\blank{1cm}\\}
% \vspace{-8pt}
\label{tab:time-to-commit-summary}
\end{table*}

We analyzed this relationship using two dependent variables: the time between opening the issue and the first/last commit, as well as four independent variables: (1) issue post sentiment, (2) issue post tone, (3) overall sentiment of issue comments, and (4) overall tone of issue comments. The two dependent variables generated the same results in terms of significant relationships with the independent variables. Table~\ref{tab:time-to-commit-summary} summarizes these relationships.

When all issues were included, Kruskal-Wallis tests indicated significant differences on both dependent variables at alpha level .001 for all independent variables except issue post sentiment. The post-hoc analysis has revealed several significant results ($p<0.001$):

\begin{itemize}[leftmargin=*]
    \item An \textit{excited} tone in issue posts was associated with longer time for issue resolution (both times until the first and the last commit) than a \textit{sad} or \textit{frustrated} tone or \textit{no tone}.
    \item A \textit{sad} tone in issue comments was associated with longer time for issue resolution than \textit{no tone}.
    \item A \textit{positive} sentiment in issue comments was associated with shorter time for issue resolution than a \textit{neutral} sentiment.
\end{itemize}

When the issues were separated into categories based on \textbf{discussion length}, the only significant differences were found in the \textit{short} and \textit{medium} discussion categories between the comment sentiments and the time to the first/last commits. Focusing on the \textbf{roles of issue posters}, in both \textit{contributors} and \textit{non-contributors} categories, the only significant relationships were again found with the comment sentiments as the independent variable. When the issues were separated by the \textbf{issue types}, no significant association was found in both \textit{feature requests} and \textit{bug reports}. When we investigate issues with different \textbf{commit sizes}, the independent variable that associated with the issue resolution times were different in the three commit size categories as summarized in Table~\ref{tab:time-to-commit-summary}.

\vspace{4pt}
\begin{tcolorbox}[colframe=black,colback=gray!10,boxrule=0.5pt,arc=.3em,boxsep=-1mm]
\textbf{Insight 4}: \textit{Positive} comment sentiments were associated with shorter resolution time, particularly in issues that needed only small commits. \textit{Sad} issue posts or comments were associated with shorter issue resolution time, particularly in issues with medium or large-sized commits.
\end{tcolorbox}

%% file: s_discussion.tex
\section{Discussion}
We would like to emphasize that in the above analysis, we do not claim causal relationships. Rather, we focused on understanding the correlational impacts of sentiments and tones expressed by the diverse OSS community members in issue tracking systems on various discussion and development measures. Overall, our study painted an extensive and complex picture of the relationships that the sentiments and tones of issue discussions have with the factors that they might have influenced.

While we used existing tools in identifying the sentiments and tones, we made considerable efforts to investigate an optimized configuration for these tools and validated the automated results. We argue that future studies using existing tools and pre-trained models for affective analysis should follow a similar approach. Moreover, our methods can inform best practices when adopting affective analysis tools. For example, we found that a sentence-level analysis with later aggregation has resulted in better performance than a post-level analysis for both IBM Tone Analyser and Senti4SD.

To our surprise, we did not find much significant relationships in the sentiments and tones between issue posts and the corresponding comments. There were, of course, cases where the issue comments expressed a certain tone in respond to the sentiments and tones of the issue post. Nevertheless, in general, people tend not to follow a pattern when reacting to the sentiments and tones expressed in the issue posts. This can be explained by the maturity of the three communities we focused on. However, further investigation is needed to understand extreme cases where a particular sentiment or tone (e.g., a \textit{frustrated} or \textit{sad} tone) could sway the tones in the follow-up comments that may impact the community confidence or health.

The sentiments and tones in the issue posts, however, did have a significant association with the discussion and development-related measures. Particularly, a \textit{negative} sentiment and a \textit{sad} tone in the posts seemed to have indicated the severity of the issue, and were associated with shorter response time, more discussions, and shorter issue resolution time. In previous studies on Jira data, however, negative sentiments were found to be associated with unresolved issues and longer issue fixing time~\cite{Ortu2015, Valdez2020}. This difference could have resulted from the different natures of Jira and GitHub issue discussions.
% and sentiment analysis tools. The tools used in these studies are SentiStrength, SentiCR, and SentiStrengthSE trained on the social web, code review comments, and Jira, respectively, while we used Senti4SD which trained on the StackOverflow similar dataset to GitHub. Besides, contributors of these ML repositories are from different fields of science. However, participants of the selected dataset in mentioned studies have more experience in computer science. Therefore, this difference might have affected how the developers and other community members reacted to the sentiments and tones embedded in the discussion. 
Further, the additional tone analysis conducted in our study provided more fine-tuned information than the previous work about the particular affective states that had a potential impact. 

The importance of sentiments and tones in the issue comments, however, were frequently overlooked in the related literature. Our analysis extends the existing literature in this area. Particularly, the direct relationship between the affective states in the comments and the issue discussion length indicates that the overall discussion's sentiments and tones can potentially reflect the complexity of the issue. Moreover, we found that the issue's codebase impact moderated the relationships of the issue comments' sentiments and tones with the issue resolution time. For example, a \textit{sad} tone in comments was associated with longer discussions only in issues with a medium commit size; similarly, \textit{positive} sentiments were associated with shorter resolution time, but mostly in issues resolved by small commits. These results complemented Sinha et al.'s finding that positive sentiments in commit messages are associated with smaller commit sizes~\cite{Sinha2016}, as well as Asri et al.'s finding that negative sentiments in code reviews took more time to complete~\cite{Ikram2019}.

Overall, our results about the relationships of the sentiments and tones expressed in the issue discussions with various discussion and development-related measures can support OSS community members in making and moderating effective discussions. For example, a \textit{negative} sentiment and a \textit{sad} tone in the issue post may indicate that the issue needs immediate attention and potentially complicated to address. Moreover, automated tools can be developed by leveraging the knowledge about the impacts of sentiments and tones in issue discussions. For example, sentiments and tones in issue discussions, as well as the potential impacts, can be made explicit to help OSS contributors anticipate the severity and complexity of the issue or identify and prevent potentially toxic discussions. Designing these types of tools would be our next steps.

%% file: s_conclusion.tex
\section{Threats to validity}
First, although the analyzed projects were carefully selected and resulted in a relatively large dataset, we were only able to focus on three OSS projects due to the significant data collection, validation, and analysis efforts. Future study with other types of OSS projects is needed to evaluate generalizability of our results. Second, the automated tools for detecting the sentiments and tones may not report accurate results. To this end, we have made an extensive investigation to validate the automated results and ensure an optimal configuration of the tools on our dataset. Still, the remaining inaccuracy of the tools may affect the validity of our analysis results. Third, when analyzing the issues linked to commits, we relied on links manually added by OSS contributors. While we cannot claim that all the issues led to a commit were identified, the dataset used is sufficiently big for this part of the analysis. Fourth, we only focused on the overall affective states of issue comments, aggregated across all comments for each issue; we also did not separately consider comments made before and after issue closure. This approach can miss the nuance changes of sentiments and tones within issue discussion threads and overlook the impacts of each comment on the rest of the discussion, particularly for long issue discussions. Exploring these aspects can be interesting future work. 
%Fifth, in this research, we have not considered separating comments before and after issue closure. It can be named one of our limitations in checking their numbers and the impact of comments on the discussion after the issues got closed.  Hence,  we can investigate these comments' affective states and their significance on our results for our future study. Sixth, our goal in this study was to investigate the statistical significance that exists in these repositories. So, it was too much attention to statistics instead of explaining the strength of these types of relations, which can be introduced as our limitation. 
Finally, we were only able to identify correlational relationships in this study. Future work such as a user study involving OSS contributors is needed to investigate the causal impacts of sentiments and tones in issue discussions.

\section{Conclusion}
In this paper, we reported on an extended empirical study aiming to understand the correlational impacts of sentiments and tones in issue discussions. Through the collection, validation, and analysis of a dataset comprising three large OSS projects, we have identified various discussion and development-related measures that were associated with the sentiments and tones in the issue posts and comments. Our effort will help draw a comprehensive picture of the role of sentiments and tones in issue discussion with respect to the OSS project and community. Insights gained from our findings also have practical implications for supporting OSS community members and designers of OSS tools to better support community engagement.